\documentstyle[epsf,psfig,aps,multicol,amsmath,amsfonts,amssymb]{revtex}
\draft
\begin{document}
\title{Arbitrary qubit rotation through nonadiabatic evolution}
\author{Asoka Biswas and G. S. Agarwal}
\address{Physical Research Laboratory, Navrangpura, Ahmedabad-380 009, India}
\date{\today}
\maketitle
\begin{abstract}
We show how one can perform arbitrary rotation of any qubit, using delayed laser
 pulses  through nonadiabatic evolution, i.e., via transitions among the
adiabatic states. We use a double-$\Lambda$ 
scheme and use a set of control parameters such as
detuning, ratio of pulse amplitudes, time separation of the two pulses for realizing
different rotations of the qubit. We also investigate
 the effect of different kinds of chirping, namely linear chirping and 
hyperbolic tangent chirping. 
 Our work using nonadiabatic evolution
adds to the flexibility in the implementation of logic gate operations and shows
how to achieve control of quantum systems by using different types of pulses.
\end{abstract}
\pacs{PACS No(s).: 03.67.-a, 32.80.-t}
\begin{multicols}{2}
\section{Introduction}
The growing field of quantum computation and quantum information involves logical operations with 
quantum states. The single qubit operation is its inversion and 
rotation whereas the controlled-NOT and phase-gates  require two qubits. 
Any specific computation consists of proper manipulation using these 
operations \cite{qinf}.

All these can be implemented on the basis of one's ability to prepare the 
atomic system in desired states. Such a preparation has been studied using a variety of techniques using external fields.
 For example, by Rabi flopping \cite{haroche}
 one can transfer the population completely to another orthogonal state. 
However this requires perfect control on pulse area.
A more efficient way to transfer populations has been proposed by  Oreg {\it et al.\/} \cite{oreg}. 
They have shown how to transfer the population between the
dipole forbidden levels in multilevel systems using the idea of {\it adiabatic following} \cite{allen}.
They have predicted the use of two delayed pulses in a three level $\Lambda$ system 
in the {\it large  detuning} domain for this purpose.  
Their idea of counter-intuitively ordered pulses has been 
implemented in a technique of population transfer called Stimulated Raman 
Adiabatic Passage 
(STIRAP) \cite{stirap} 
. Under conditions of {\it adiabaticity},
the atom follows the evolution of one of the eigenvectors of the time-dependent 
Hamiltonian. Grobe {\it et al.\/}, under the same condition of adiabaticity,
used the idea of  delayed pulses in counterintuitive
sequence in the context of formation of shape-preserving
pulses (adiabatons) \cite{adiabaton}. Note that, in STIRAP, both the 
pulses can in principle be kept  {\it resonant} with the corresponding atomic transitions. Vitanov and 
Stenholm \cite{vita} investigated the effect of Raman detuning on the final 
state population in STIRAP starting from one of the two lower  states. They 
used two 
delayed Gaussian pulses for this purpose. They found that, 
as the detuning increases, the final state population decreases. Sola 
{\it et al. \/} \cite{sola}  used two contemporary chirped pulses  
and  investigated how the final state population varies with the 
chirp rate of the pulses. They demonstrated the role of detuning in 
minimizing  the  intermediate level population.

All the above references discuss the population transfer between the 
unperturbed states of the atoms.  Chang {\it et al\/}. \cite{chang} found
a way to produce a final state which was a superposition of the bare atomic states.
They  used two contemporary chirped pulses. They have shown how the selectivity
 of the final states depends upon the relative sign of chirping rate and 
detuning.  Further, 
Unanyan {\it et al\/}. in the tripod-STIRAP scheme\cite{shore} have shown how one can create an arbitrary superposition of two atomic states using a sequence of three pulses in a four level system 
and how the superposition can be controlled by changing the relative delay 
between the pulses. Marte {\it et al\/}. \cite{marte} and more recently Vitanov 
{\it et al\/}. \cite{frac} suggested another method, called fractional STIRAP, in which, 
the two counterintuitive pulses terminate at the same time to produce appropriate
superposition states. In a recent development, Kr\'al and Shapiro 
\cite{shapiro1} has found another unique way of creating a superposition of 
many energy-eigenstates by the method of shaped adiabatic passage (SAP). 

So far we have discussed different methods of either the population transfer 
or the preparation of superpositions, all starting from a given unperturbed state. 
 Now the question arises whether one can create a 
superposition, starting from a superposition state, i.e., produce essentially
the rotation of a qubit. 
Renzoni and Stenholm \cite{renzoni} used four laser pulses which coupled four
ground levels to a common excited level and created a superposition of two
ground states, starting from a superposition of {\it two other ground states}.
Most recently Kis and Renzoni \cite{kis} have discovered how to create the
superposition {\it in the same basis} as of the initial superposition state.
They have used the tripod-STIRAP scheme to show how the relative phase of
the control pulse controls the superposition of the atomic states in the final
superposition. This is quite relevant in the context of qubit
rotation. If one thinks of the initial superposition state as one qubit state,
then creating another superposition in the same basis refers to the rotation
of the qubit in the Hilbert space of those two basis states.

All the above methods are based on the adiabatic evolution of atomic
system. The question arises, whether one can go beyond the adiabatic limit.
In the present paper, we explore this possibility in the context of  a three
level $\Lambda$-system interacting with two laser pulses.
 In nonadiabatic evolution, transition occurs between the different 
adiabatic states of the system \cite{eberly}. 
 During this rotation the state no longer remains confined in the 
two-dimensional Hilbert space of the basis states comprising the superposition. 
Thus a different superposition of the basis states is prepared in the long
time limit.

We use a double-$\Lambda$ configuration for the demonstration of the rotation of
a qubit. We note that the light propagation through an atomic medium in such a
configuration has been studied \cite{double,huss}.  

The structure of the paper is as follows. In Sec. II, we describe the model
configuration and relevant equations for the process of qubit rotation. In Sec. III, we present detailed
 numerical results on the rotation of a qubit. We discuss the role of 
detuning and other pulse parameters in this context. In Sec. IV, we present a
comparison of our results with those obtained using two-level approximation. In Sec. V,
we discuss the nonadiabaticity in the evolution. In Sec. VI, we consider 
the effect of chirped pulses on the  rotation of a qubit. It may be noted that
pulses with appropriate time-dependent phases have been used to achieve 
coherent control of quantum systems \cite{cohcont}. In the Appendix, 
we show how the traditional STIRAP can be used to prepare the atomic system in a state which is orthogonal to the initial superposition.

\section{Model for qubit rotation}
Consider the four level configuration as shown in Fig.\ \ref{config}. The 
levels $|g\rangle$ and $|f\rangle$ are coupled to the upper excited levels 
$|e\rangle$ and $|1\rangle$ by input pulses. We will show how specific 
choices of these pulses help in preparing some initial superposition of 
$|g\rangle$ and $|f\rangle$ and how this superposition can be rotated 
with the help of two delayed pulses. Note that the basis states of the qubit, 
under consideration, are $|g\rangle$ and $|f\rangle$. The states $|g\rangle$ 
and $|f\rangle$ are long-lived states and that the initial and final states should
not have any population in the excited states $|e\rangle$ and $|1\rangle$.

{\narrowtext
\begin{figure}
\epsfxsize 8cm
\centerline{
\epsfbox{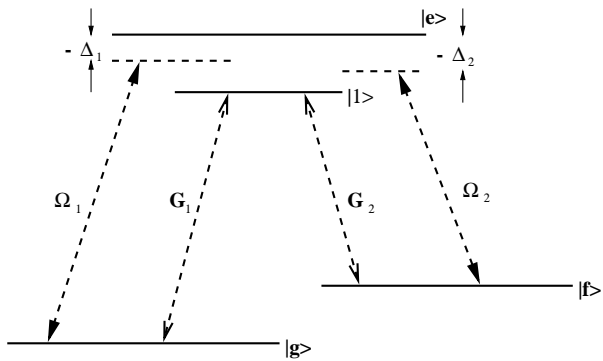}}
\caption{Level configuration for a double-$\Lambda$ system. Here $G_1$ and $G_2$
are one-half of the Rabi frequencies of the applied classical fields which couple 
the level
$|1\rangle$ to $|g\rangle$ and $|f\rangle$, respectively. The classical fields 
with  Rabi frequencies $2\Omega_1(t)$ and $2\Omega_2(t)$
couple the level $|e\rangle$ to $|g\rangle$ and $|f\rangle$.  $\Delta_1$ and
$\Delta_2$ are their respective one-photon detunings.}
\label{config}
\end{figure}}

\subsection{Preparation of the initial superposition state}
We start with the atom initially in the level $|g\rangle$. We apply two cw 
fields with constant Rabi frequencies $2G_1$ and $2G_2$ which couple the state 
$|1\rangle$ with $|g\rangle$ and $|f\rangle$, respectively. We assume these 
fields to be resonant with the corresponding transitions.
When the cw fields are switched on, the system is prepared in  the state 
\begin{equation}
|i\rangle=\alpha |g\rangle + \beta e^{i\phi} |f\rangle,
\label{initial}
\end{equation}
where, $\alpha=G_1/G,~\beta=G_2/G,~~G=\sqrt{G_1^2+G_2^2}$ and $\phi$ is the relative phase between $|g\rangle$ and $|f\rangle$.
This is the desired state, which will be considered as the initial state of the qubit in the next phase of evolution.
   
\subsection{Basic equations for qubit rotation}
We will now discuss how one can rotate the qubit in state (\ref{initial}), i.e.,
how one can create another superposition of $|g\rangle$ and $|f\rangle$. When
the cw fields are switched off, we apply two chirped laser pulses with 
electromagnetic fields given by
\begin{mathletters}
\begin{eqnarray}
E_1(t)&=&\varepsilon_1(t) e^{-i\omega_1 t} e^{-i\phi_1(t)} +\textrm{c.c.},\\
E_2(t)&=&\varepsilon_2(t) e^{-i\omega_2 t} e^{-i\phi_2(t)} +\textrm{c.c.},
\end{eqnarray}
\end{mathletters}
where, $\varepsilon_{1,2}(t)$ are the envelopes of the pulses, $\omega_{1,2}$ are the 
carrier frequencies, $\dot{\phi}_i(t)=\chi_iF_i(t)$ $(i=1,2)$ corresponds to the applied
chirping with chirping parameter $\chi_{1,2}$ and time-dependence of $F_i(t)$ represents 
the chirping profile. These pulses couple the level $|e\rangle$ with $|g\rangle$ and $|f\rangle$, respectively.

The interaction Hamiltonian in the interaction picture and in the rotating wave approximation  is 
\begin{eqnarray}
H&=&\hbar \left[ \Omega_1(t) e^{-i\Delta_1 t}e^{-i\phi_1(t)} |e\rangle\langle g|+\Omega_2(t)e^{-i\Delta_2 t}e^{-i\phi_2(t)}|e\rangle\langle f|\right.\nonumber\\
&&\hspace*{3cm}\left.+\textrm{h.c.}\right],
\label{hamil}
\end{eqnarray}
where, $2\Omega_1(t)=2d_{eg}\varepsilon_1(t)/\hbar$ and $2\Omega_2(t)=2d_{ef}\varepsilon_2(t)/\hbar$ are the time-dependent Rabi frequencies, $d_{ej}$ $(j=g,f)$  are the dipole matrix elements
of transitions $|e\rangle \leftrightarrow |j\rangle$, $\Delta_1=\omega_1-\omega_{eg}$, $\Delta_2=\omega_2-\omega_{ef}$ are the respective one-photon detunings of the fields, and $\omega_{ej}$ are the frequencies of the atomic transitions $|e\rangle \leftrightarrow |j\rangle$. Here we choose the envelopes $\Omega_1(t)$ and $\Omega_2(t)$ as
\begin{mathletters}
\begin{eqnarray}
\Omega_1(t)&=&\Omega_{01}\exp{[-(t-T)^2/\tau^2]}e^{-i\delta},\\
\Omega_2(t)&=&\Omega_{02}\exp{(-t^2/\tau^2)},
\end{eqnarray}
\label{pulses}
\end{mathletters}
where, $\Omega_{0j}$ $(j=1,2)$ is one-half of the peak Rabi frequency, $\tau$ is the
half-width of both the pulses, $T$ is their relative time separation, and $\delta$ is the relative phase difference of the pulses. 
The wave function of the atomic system can be expanded in the basis states 
$(|e\rangle, |g\rangle, |f\rangle)$ as
\begin{equation}
|\Psi (t)\rangle = c_e(t)|e\rangle + c_g(t)|g\rangle + c_f(t)|f\rangle,
\end{equation}
where, $c_j(t)$ $(j=e,g,f)$ are the probability amplitudes for the corresponding
states. 
Using Eq.\ (\ref{hamil}) and the following transformations of the amplitudes
\begin{eqnarray}
d_e&=&c_ee^{i\Delta_1 t}e^{i\phi_1(t)},\nonumber\\
d_g&=&c_g,\\
d_f&=&c_fe^{i(\Delta_1-\Delta_2)t}e^{i[\phi_1(t)-\phi_2(t)]},\nonumber
\end{eqnarray}
we arrive at the following time dependent equations for the $d_j(t)$'s:
\begin{mathletters}
\begin{eqnarray}
\dot{d}_e&=&i(\Delta_1+\dot{\phi}_1)d_e-i[\Omega_1(t)d_g+\Omega_2(t)d_f],\\
\dot{d}_g&=&-i\Omega_1^*(t)d_e,\\
\dot{d}_f&=&i[\Delta_1-\Delta_2+(\dot{\phi}_1-\dot{\phi}_2)]d_f-i\Omega_2^*(t)d_e.
\end{eqnarray}
\label{dotequ}
\end{mathletters}
These equations can be solved numerically under the initial conditions [Eq.\ (\ref{initial})]
\begin{equation}
\label{condi}
d_g(-\infty)=\alpha, ~~~~~ d_f(-\infty)=\beta e^{i\phi}.
\end{equation}

{\narrowtext
\begin{figure}
\epsfxsize 8cm
\centerline{
\epsfbox{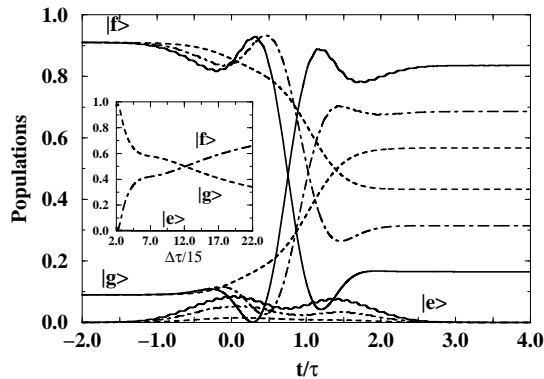}}
\caption{Effect of the detuning parameter $\Delta$  on the
time dependence of the populations of the levels $|e\rangle$, $|g\rangle$, 
and $|f\rangle$
for different values of   $\Delta\tau=45$ (solid curve), $\Delta\tau=60$ 
(dot-dashed curve), and $\Delta\tau=120$ (dashed curve). The other parameters 
 are $\alpha=0.3$, $\phi=\pi/2$, $\Omega_{01}\tau=\Omega_{02}\tau=15$,
$\delta=0$, $\chi_1=\chi_2=0$, and $T=4\tau/3$.
 The variation of populations of the levels $|g\rangle$ (dashed curve) and
$|f\rangle$ (dot-dashed curve) with $\Delta$ at long time $(t=15\tau)$ is shown in the inset with the same numerical parameters as above.} 
\label{detune}
\end{figure}}

{\narrowtext
\begin{figure}
\epsfxsize 8cm
\centerline{\epsfbox{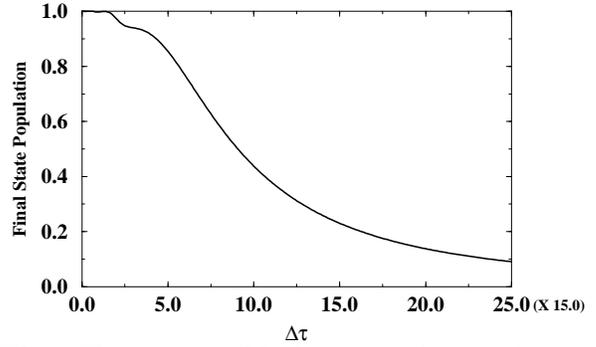}}
\caption{The variation of final state population at long time with $\Delta$ for 
$\alpha=1$. Other parameters are the same as in Fig.\ \ref{detune}.}
\label{detuneB}
\end{figure}}

\section{arbitrary qubit rotation controlled by different pulse parameters}

\subsection{Effect of one-photon detuning}
We assume that the two delayed pulses (\ref{pulses}) are equally detuned from the corresponding transitions, 
i.e., $\Delta_1=\Delta_2=\Delta$, say. This means that both the pulses are 
one-photon detuned, but satisfy the condition of two-photon resonance. We also
 assume that they are unchirped, i.e.,
$\chi_1=\chi_2=0$. We further assume that, the pulses do not have any relative phase difference, i.e., $\delta=0$ and have
equal magnitudes, i.e., $\Omega_{01}=\Omega_{02}$. We start with the initial 
atomic superposition (\ref{initial}) choosing $\alpha=0.3$ and $\phi=\pi/2$.
We integrate Eqs.\ (\ref{dotequ}) to obtain the time-dependence of the various 
amplitudes.

{\narrowtext
\begin{figure}
\epsfxsize 8cm
\centerline{
\epsfbox{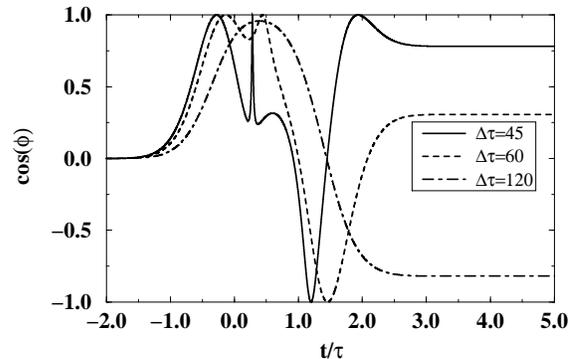}}
\caption{Temporal variation of the relative phase  between
the levels $|g\rangle$ and $|f\rangle$ for different values of $\Delta$ is
shown. The other
parameters are the same as in Fig.\ \ref{detune}. Note that the phase
becomes constant in the long time limit.}
\label{relphase}
\end{figure}}

We plot the variation of populations in the levels $|e\rangle$, $|g\rangle$, and
$|f\rangle$ with time for different values of $\Delta$, in Fig.\ \ref{detune}.
We note that in the long time limit the population gets redistributed in the 
levels $|g\rangle$ and $|f\rangle$ with almost no population in $|e\rangle$. As
the value of $\Delta$ varies, the populations in $|g\rangle$ and $|f\rangle$
also vary. We show in the inset of Fig.\ \ref{detune} how the populations in 
$|g\rangle$ and $|f\rangle$ in the long time limit vary with $\Delta$. In this
limit, the intermediate level population becomes almost zero. This is the case even when $\alpha=1$, i.e., if the initial state is
$|g\rangle$. Fig.\ \ref{detuneB} clearly shows remarkably large change in the 
final state population with $\Delta$ (see also \cite{vita}). This, as we will
see in Sec. V, has bearing on the adiabatic behavior.

  The relative phase $\phi$ of the levels $|g\rangle$ and $|f\rangle$
is given by 
\begin{equation}
\cos \phi(t) = \textrm{Re}[c_g^*c_f/|c_g|.|c_f|].
\end{equation}
In Fig.\ \ref{relphase}, we have shown the time-dependence of $\cos \phi(t)$. 
Clearly, the phase at the long time limit becomes constant. This means, a  
different {\it coherent} superposition of $|g\rangle$ and $|f\rangle$ has been 
created. This implies to a rotation of the concerned qubit, which is 
controlled by one-photon detuning $\Delta$.

Note that for intermediate times, the level $|e\rangle$ gets populated.
Thus  the atomic system initially in
the state $|i\rangle$ no longer remains confined in the two-dimensional 
Hilbert plane of $|g\rangle$ and $|f\rangle$.
Rather, it travels through the three-dimensional Hilbert 
space of the levels $|e\rangle$, $|g\rangle$, and $|f\rangle$ and finally comes
back to the initial space of states $|g\rangle$ and $|f\rangle$, but with 
a {\it different composition} of them. 

From the above discussions, one can infer that the amount of rotation can 
be suitably controlled by the one-photon
detuning parameter $\Delta$. For example, for $\Delta\tau\sim 180$, one 
can prepare an equal superposition state of the levels $|g\rangle$ and $|f\rangle$ with phase $\phi=2\pi/3$ (see inset, Fig.\ \ref{detune}).
We emphasize that for $\Delta=0$, the level $|e\rangle$ remains populated 
at large times. So large values of the detuning $\Delta$ are needed for proper qubit
rotation. 

It is  interesting to note that the time separation $T$ between the pulses
 affects the qubit rotation as shown in Fig.\ \ref{time} which depicts the 
variation of 
populations in $|g\rangle$ and $|f\rangle$ with $T$. It should
be noted that at long time, the level $|e\rangle$ does not get populated.

{\narrowtext
\begin{figure}
\epsfxsize 8cm
\centerline{
\epsfbox{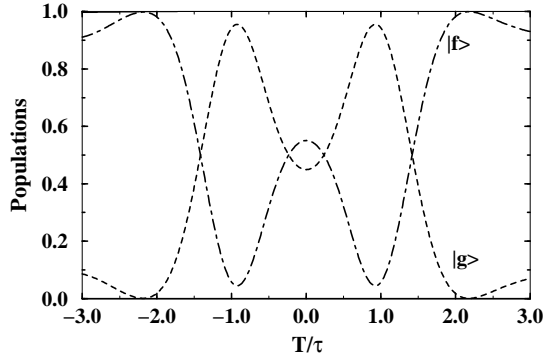}}
\caption{Variation of the populations in the levels 
$|g\rangle$ and $|f\rangle$ in the long time limit 
($t=15\tau$) with the time separation $T$ of the pulses, for $\Delta\tau=75$. 
The other parameters used are the same as in Fig.\ \ref{detune}.}
\label{time}
\end{figure}}

It should be mentioned here that one can get approximated analytical solutions 
of Eqs.\ (\ref{dotequ}), putting $\chi_i=0$ $(i=1,2)$ \cite{vita,hioe,isq}. However, all these solutions
are for $\alpha=1$. For $\alpha\neq 1$, as discussed in the present section, it
is very difficult to  obtain the solution analytically.

{\narrowtext
\begin{figure}
\epsfxsize 8cm
\centerline{
\epsfbox{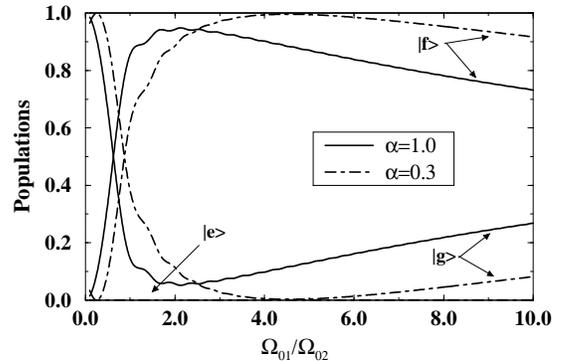}}
\caption{Variation of populations in the levels $|e\rangle$, $|g\rangle$
and $|f\rangle$ with the ratio of the amplitudes of the Rabi frequencies
$\Omega_{01}/\Omega_{02}$ at long time $(t=15\tau)$. The 
numerical parameters used are $\phi=\pi/2$, $\Omega_{02}\tau=15$, $\delta=0$, $\chi_1=\chi_2=\chi=0$, and $\Delta \tau=75$.}
\label{ampliA}
\end{figure}}

{\narrowtext
\begin{figure}
\epsfxsize 8cm
\centerline{\epsfbox{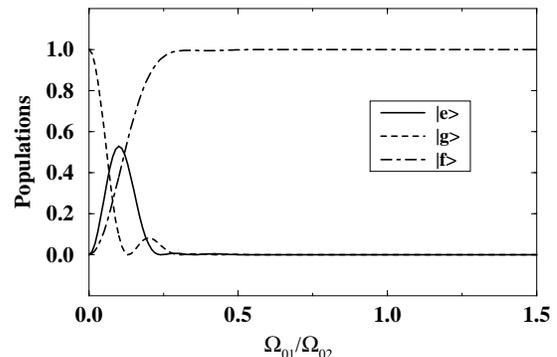}}
\caption{Variation of populations in long time limit ($t=15\tau$) with the ratio $\Omega_{01}/\Omega_{02}$ for $\alpha=1$ and
$\Delta\tau=0$ is shown. All
the other parameters are the same as in Fig.\ \ref{ampliA}.}
\label{ampliB}
\end{figure}} 

\subsection{Effect of relative amplitudes of the pulses}
The relative amplitude of the two pulses is expected to determine the final state 
of the atom.  We continue
to keep both the pulses equally detuned by an amount of $\Delta$.  
In Fig.\ \ref{ampliA}, we show the long time behavior of
populations in the levels $|e\rangle$, $|g\rangle$, and $|f\rangle$ 
as functions of the parameter $\Omega_{01}/\Omega_{02}$. We present results
 for both the cases when the atom starts in the  state $|g\rangle$ and the
state $|i\rangle$ with $\alpha=0.3$ and $\phi=\pi/2$.
Note that the population in the level $|e\rangle$ is negligible for large times.
The population is redistributed in the levels $|g\rangle$ and $|f\rangle$. Thus,
a different superposition is created at the end of evolution. The composition 
varies if one changes the ratio $\Omega_{01}/\Omega_{02}$. For example, 
for $\Omega_{01}/\Omega_{02}=1$, one gets an equal superposition of $|g\rangle$
and $|f\rangle$, for $\alpha=0.3$. It is thus clear that  the ratio
of the amplitudes of the two pulses can serve as a control parameter for
producing rotation of a  qubit. Note that for $\alpha=1$ and $\Delta =0$ only 
orthogonal rotation of the qubit is possible for all values of $\Omega_{0i}$ 
[see Fig.\ \ref{ampliB}].

{\narrowtext
\begin{figure}
\epsfxsize 8cm
\centerline{
\epsfbox{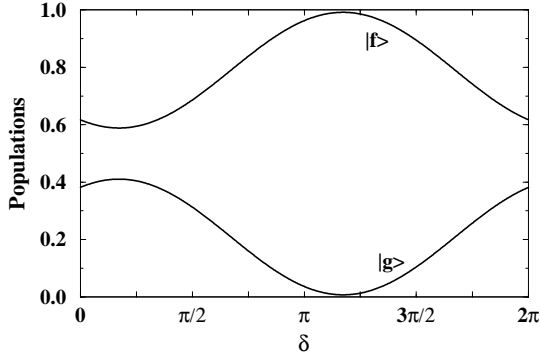}}
\caption{Variation of the population of the levels $|g\rangle$ and
$|f\rangle$ with the relative phase $\delta$ of the two applied
electric fields  at long time ($t=15\tau$) for the parameters
$\alpha=0.3$, $\phi=\pi/2$, $\Omega_{01}/\Omega_{02}=1.0$, $\Delta\tau=75$, and $\chi=0$.
}
\label{phase}
\end{figure}}

\subsection{Effect of relative phase of the two pulses}
We will now discuss how the amount of qubit rotation can be controlled by the 
relative phase $\delta$ between the two pulses.
We work under the two-photon resonance condition and take the pulses to have equal amplitudes.
The periodic dependence of the populations of $|g\rangle$ and $|f\rangle$ on 
$\delta$ in the long time limit for the initial condition (\ref{condi}) is
 shown in Fig.\ \ref{phase}. Thus the relative phase of the input
pulses can be used as an useful control parameter for the qubit rotation.

It should be mentioned here that, if $\alpha=1$, i.e., if the atom is initially
prepared in the state $|g\rangle$, then there is no $\delta$-dependence of the
populations in the states $|g\rangle$ and $|f\rangle$. This is because the
Eqs. (\ref{dotequ}) can be redefined with an extra complex phase term in the 
probability amplitudes. For example, the replacements $\tilde{d}_e=d_ee^{i\delta}$, $\tilde{d}_f=d_fe^{i\delta}$, and $\tilde{d}_g=d_g$ keep the form of the equations unchanged. Thus $\delta$ does not show up in  the square moduli of 
these amplitudes. However, the wave functions would be different for different
  values of $\delta$ and the probability amplitudes become different. Thus, if 
the atom is initially in the state $|g\rangle$, then a change in $\delta$ 
cannot produce arbitrary rotation of $|g\rangle$.  

\section{Comparison with the results in two-level approximation }
We write the state orthogonal to $|i\rangle$ [Eq.\ (\ref{initial})] as 
\begin{equation}
|k\rangle=\beta e^{-i\phi}|g\rangle-\alpha |f\rangle.
\label{final}
\end{equation}
When the detuning $\Delta$ of the pulses becomes much larger than the pulse amplitudes, then one can ignore the time
variation of $d_e$ \cite{oreg}. The Eqs. (\ref{dotequ}) in the ($|i\rangle, |k\rangle$) basis can be written as
\begin{equation}
\left( \begin{array}{c}
\dot{d}_i \\
\dot{d}_k 
\end{array} \right) =
-i
\left( \begin{array}{cc}
\Delta_e & \Omega_e \\
\Omega_e^* & -\Delta_e 
\end{array} \right) 
\left( \begin{array}{c}
d_i\\
d_k
\end{array} \right),
\label{twolev}
\end{equation}
where,
\begin{eqnarray}
f_1(t)&=&\alpha \Omega_1^*(t)+\beta e^{-i\phi}\Omega_2^*(t),\nonumber\\
f_2(t)&=&\beta e^{i\phi}\Omega_1^*(t)-\alpha\Omega_2^*(t),
\label{neweq}
\end{eqnarray}
and $\Omega_e=f_1f_2^*/\Delta$, $\Delta_e=(|f_1|^2-|f_2|^2)/2\Delta$.
 Thus the three-level atomic system can be approximated as a two-level system 
with effective Rabi-frequency $\Omega_e$ and effective detuning $\Delta_e$ 
[Eq.\ (\ref{twolev})]. It is well-known that in a two-level atomic system, if detuning of the pulse is much larger than the pulse amplitude, then the induced polarization of the atom 
follows the time-evolution of the population term \cite{allen}. This is referred to as adiabatic following. Similarly, if 
$|\Delta_e|\gg |\Omega_e|$, then, adiabatic following occurs in a three-level 
system also.  

{\narrowtext
\begin{figure}
\epsfxsize 8cm
\centerline{
\epsfbox{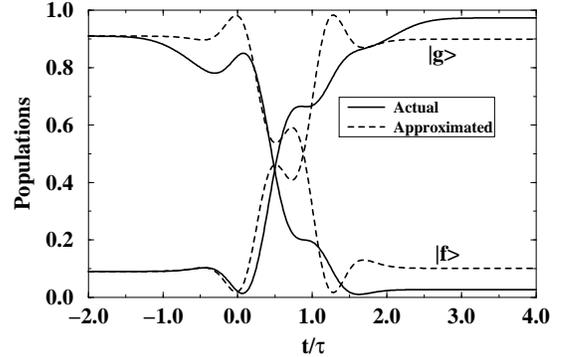}}
\caption{The variation of populations in the levels $|g\rangle$ and $|f\rangle$
as obtained from
exact numerical results and from two-level approximation for $\Delta\tau=30$. 
The other parameters are $\alpha=0.3$, $\phi=\pi/2$, $\Omega_{01}\tau=\Omega_{02}\tau=15$, $\delta=0$, $\chi_1 =\chi_2=0$, and $T=4\tau/3$.}
\label{new4a}
\end{figure}}

{\narrowtext
\begin{figure}
\epsfxsize 8cm
\centerline{\epsfbox{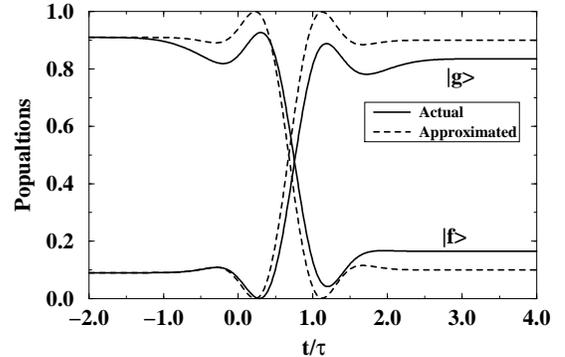}}
\caption{The variation of populations in the levels $|g\rangle$ and $|f\rangle$
as obtained from
exact numerical results and from two-level approximation for $\Delta\tau=45$. 
The other parameters are the same as in Fig.\ \ref{new4a}.}
\label{new4b}
\end{figure}}

We present the exact and two-level approximated results for the variation of the populations in $|g\rangle$ and $|f\rangle$
in Figs. \ref{new4a} and \ref{new4b} for different values of $\Delta$. Clearly, for smaller values of $\Delta$, the two-level 
approximation is not valid. But still {\it rotation of the qubit occurs}. Only when $\Delta \gg \Omega_{0i}$,
exact numerical result and two-level approximated result match, as we stated earlier. This is because, in this case, the level $|e\rangle$ is 
hardly populated during the evolution and the system thereby keeps itself confined in the Hilbert plane of $|g\rangle $ and 
$|f\rangle$.  Thus the rotation of a qubit can be obtained even for values of
$\Delta$ for which two-level approximation is not valid.
 
{\narrowtext
\begin{figure}
\epsfxsize 8cm
\centerline{
\epsfbox{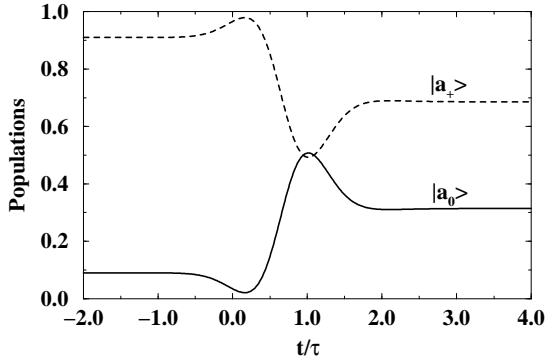}}
\caption{Variation of populations in the adiabatic states $|a_0\rangle$ and $|a_+\rangle$ [Eqs.\ (\ref{adiabatic})] with time
for $\Delta\tau =60$. The other parameters are same as in Fig.\ \ref{new4a}.}
\label{explain}
\end{figure}}

\section{Transitions among adiabatic states - Nonadiabaticity}
We further address the question of nonadiabaticity. The evolution of a quantum system is called adiabatic, if the populations of 
the adiabatic states (which are the eigenstates of the time-dependent 
Hamiltonian) remain constant during the evolution. This condition requires 
that the nonadiabatic coupling between the adiabatic states is much smaller 
than their energy difference \cite{messiah,born}. To attain this, the Hamiltonian should be 
smoothly varying with time. In a three level $\Lambda$ system, if one uses 
two smoothly varying pulses with large amplitudes in counterintuitive 
sequence, then the complete transfer of population occurs between the two 
lower lying levels of the system and also the evolution of the system becomes 
adiabatic \cite{stirap}. One has to use the pulses in Raman resonance. Even when the pulses are one-photon resonant, then for large enough 
values of the pulse amplitudes the evolution becomes adiabatic. In this light we examine the nonadiabaticity of the process
of rotation described here.
 The Hamiltonian (under the conditions $\Delta_1=\Delta_2=\Delta$ and $\chi_1=\chi_2=0$)  can be written in the ($|e\rangle$, $|g\rangle$, $|f\rangle$) basis as 
\begin{equation}
H=\left(\begin{array}{ccc}-\Delta & \Omega_1 & \Omega_2\\
\Omega_1^* & 0 & 0\\
\Omega_2^* & 0 & 0
\end{array}\right).
\end{equation}
The eigenstates (or the adiabatic states) of $H$ are given by,
\begin{mathletters}
\begin{eqnarray}
\label{adiaa}|a_0\rangle &=& \cos{\Theta}|g\rangle -\sin{\Theta}|f\rangle,\\
|a_+\rangle &=& \sin{\Theta}\cos{\Phi}|g\rangle -\sin{\Phi}|e\rangle +\cos{\Theta}\cos{\Phi}|f\rangle,\\
|a_-\rangle &=& \sin{\Theta}\sin{\Phi}|g\rangle +\cos{\Phi}|e\rangle +\cos{\Theta}\sin{\Phi}|f\rangle,
\end{eqnarray}
\label{adiabatic}
\end{mathletters}
where 
\begin{equation}
\tan\Theta=\Omega_1(t)/\Omega_2(t),
\end{equation}
 and 
\begin{equation}
\tan\Phi=\frac{2\sqrt{|\Omega_1|^2+|\Omega_2|^2}}{\Delta+\sqrt{\Delta^2+4(|\Omega_1|^2+|\Omega_2|^2)}}.
\end{equation}
The state (\ref{adiaa}) is the well-known coherent population trapping state.
It is independent of the two-photon detuning.

We have shown the temporal variation of the populations in these adiabatic states in Fig.\ \ref{explain}. 
It is clear that during the course of evolution, transfer of population occurs 
between the states, namely $|a_0\rangle$ and $|a_+\rangle$. Thus the nonadiabatic 
coupling is no more small enough to keep the process adiabatic. Clearly the system 
evolves nonadiabatically in the chosen parameter domain. 

Note that even when $\alpha=1$, for larger $\Delta$ the evolution becomes nonadiabatic [Fig.\ \ref{detuneB}] for a 
constant value of $\Omega_{0i}$. If the system were to stay in the adiabatic 
state defined by $H(t)|a_0\rangle=0$, then the final state population would be
independent of the detuning. The $\Delta$-dependence of the population in
Fig.\ \ref{detuneB} implies that the dynamics of the system does not remain
confined to the adiabatic state $|a_0\rangle$ only. Further note that for 
a given value 
of $\Delta$, large values of $\Omega_{0i}$ are required
to make the evolution adiabatic. Fig.\ \ref{ampliB} supports the adiabatic
result at large $\Omega_{0i}$ ($\Omega_{01}/\Omega_{02} \gtrsim 0.3$) for 
$\Delta=0$, i.e., the system remains essentially in the state (\ref{adiaa}).

\section{qubit rotation controlled by chirping}

So far we have discussed how the one-photon detuning of the pulses  can be used
to produce rotation of the qubit. We now consider use of the chirped pulses
for obtaining arbitrary qubit rotation.
We also examine the simultaneous effects of chirping and one-photon detuning.

{\narrowtext
\begin{figure}
\epsfxsize 8cm
\centerline{
\epsfbox{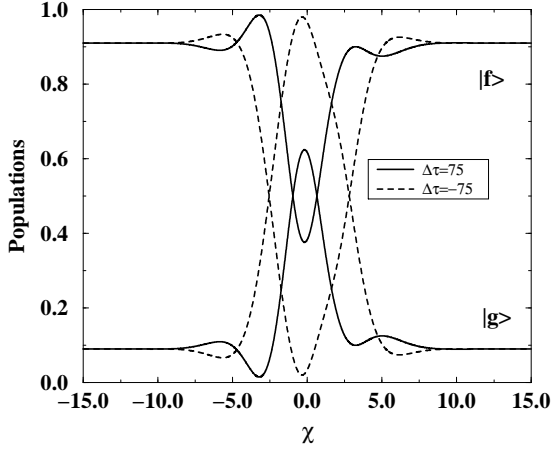}}
\caption{The variation of populations of $|g\rangle$ and $|f\rangle$ in the long time 
limit with $\chi$ (in case of linear chirping) for $\Delta\tau=75$ and $\Delta\tau=-75$ is shown. The other 
parameters are $\alpha=0.3$, $\phi=\pi/2$, $\Omega_{01}/\Omega_{02}=1$,and $\delta=0$.}
\label{linchirpA}
\end{figure}}

{\narrowtext
\begin{figure}
\epsfxsize 8cm
\centerline{\epsfbox{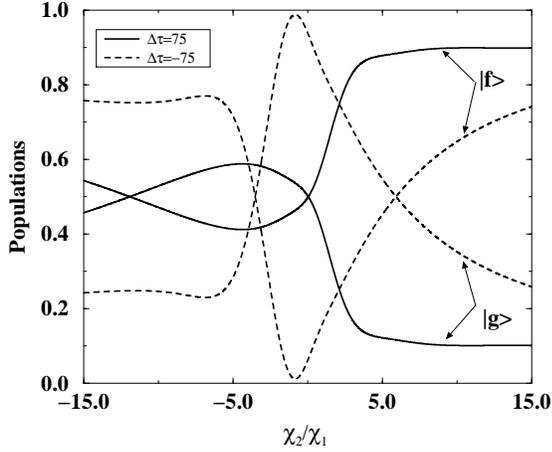}}
\caption{Variation of populations of $|g\rangle$ and $|f\rangle$ in long time limit with 
$\chi_2/\chi_1$ (taking $\chi_1=1$) is shown for $\Delta\tau=75$ and $\Delta\tau=-75$. The other 
parameters are the same as in Fig.\ \ref{linchirpA}.}
\label{linchirpB}
\end{figure}}

\subsection{Effect of linear chirping}
We assume that the pulses are linearly chirped, i.e.,
\begin{eqnarray}
\left[\frac{d}{d\left(\frac{t}{\tau}\right)}\right] \phi_1 &= &\chi_1\left(\frac{t-T}{\tau}\right),\nonumber\\
\left[\frac{d}{d\left(\frac{t}{\tau}\right)}\right] \phi_2 &= &\chi_2\left(\frac{t}{\tau}\right).
\end{eqnarray}
We further assume that $\chi_1=\chi_2=\chi$. We put $\delta=0$, $\alpha=0.3$, 
and $\Omega_{01}/\Omega_{02}=1$. We have seen from the temporal variation of
populations in the levels $|e\rangle$, $|g\rangle$, and $|f\rangle$ (results
not shown) that populations in $|g\rangle$ and $|f\rangle$ in the long time
limit becomes constant at a value depending on $\chi$ whereas the level 
$|e\rangle$ remains unpopulated. Therefore by changing the chirping rate $\chi$,
we produce arbitrary rotation to the state $|i\rangle$. In Fig.\ \ref{linchirpA}
we show the variations of populations of $|g\rangle$ and $|f\rangle$ in the 
long time limit with the chirping parameter $\chi$ for different values of $\Delta$. Clearly, we can have a rotation irrespective of the relative sign of
$\chi$ and $\Delta$, but the amount of rotation is quite dependent on it. From
Fig.\ \ref{linchirpA}, it is also clear that for a constant value of $\chi$,
one can have different rotation for different values of detuning parameter $\Delta$. Thus by varying either $\Delta$ [or $\chi$] while keeping $\chi$ [or $\Delta$]
constant, one has a way to achieve arbitrary rotation to the qubit state $|i\rangle$.

We have further found that making $\chi_1\neq\chi_2$ can also 
control the rotation. In Fig.\ \ref{linchirpB}, we have shown the variation of populations
in the levels  $|g\rangle$ and $|f\rangle$ at long time limit
with the ratio $\chi_2/\chi_1$ for two different values of $\Delta$. Note that
the level $|e\rangle$ remains unpopulated. Thus either by varying the chirping
parameter $\chi$ $(\chi_1=\chi_2=\chi)$ or by changing the ratio $\chi_2/\chi_1$
one can achieve arbitrary rotation of a qubit.

We have checked that the populations in the adiabatic states does not remain 
constant during the evolution and hence the process becomes
nonadiabatic in this case also.

{\narrowtext
\begin{figure}
\epsfxsize 8cm
\centerline{
\epsfbox{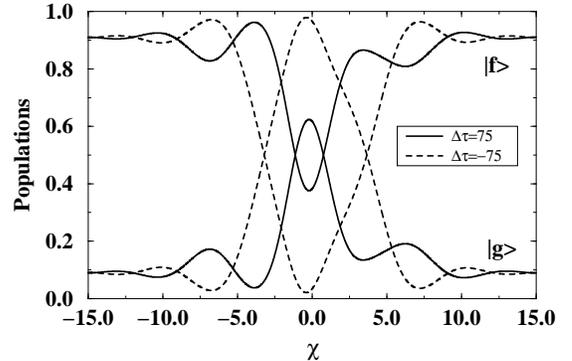}}
\caption{Variation of populations in the levels $|g\rangle$ and $|f\rangle$ in long 
time limit with $\chi$ (in case of tanh chirping) is shown for 
positive and negative values of detunings. The other parameters are same as in Fig.\ \ref{linchirpA}.}
\label{tanhchirpA}
\end{figure}}

{\narrowtext 
\begin{figure}
\epsfxsize 8cm
\centerline{\epsfbox{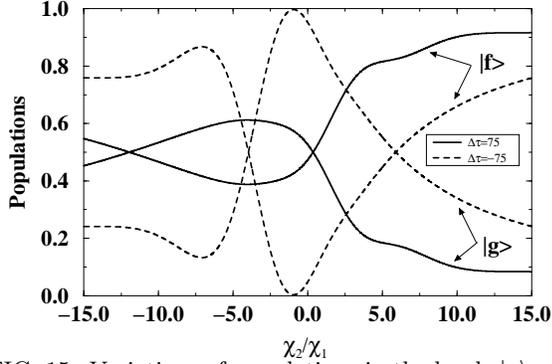}}
\caption{Variations of populations in the levels $|g\rangle$ and $|f\rangle$ in the long time limit 
with $\chi_2/\chi_1$ ($\chi_1=1$), is shown for positive and negative values of detunings. The other 
parameters are same as in Fig.\ \ref{linchirpA}.}
\label{tanhchirpB}
\end{figure}}
\subsection{Effect of hyperbolic tangent chirping}

We now assume that the chirping profile of the pulses are given by \cite{allen1}
\begin{eqnarray}
\left[\frac{d}{d\left(\frac{t}{\tau}\right)}\right] \phi_1 &= &\chi_1\tanh\left(\frac{t-T}{\tau}\right),\nonumber\\
\left[\frac{d}{d\left(\frac{t}{\tau}\right)}\right] \phi_2 &= &\chi_2\tanh\left(\frac{t}{\tau}\right).
\end{eqnarray}
In Figs.\ \ref{tanhchirpA} and \ref{tanhchirpB} we show how populations in $|g\rangle$ and $|f\rangle$ in the long time limit vary with the chirping parameters. We show the results
for equal ($\chi_1=\chi_2=\chi$) as well as for unequal ($\chi_1\neq\chi_2$)
chirping. Clearly, in this case also, rotation occurs irrespective of relative
sign of $\chi_i$'s and $\Delta$. However, if relative sign changes, amount of rotation
also changes.
 
We note that, for $\Delta=0$, both kinds of chirping do not provide any 
rotation to a general superposition state of the qubit $(\alpha\neq 1)$. 
However, if the qubit is initially in state $|g\rangle$ $(\alpha=1)$, then
both profiles provide orthogonal rotation to the state $|g\rangle$ irrespective of 
relative values of $\chi_i$'s. This is because the population in the excited 
state in the long time limit deceases as $\alpha$ increases to unity.

\section{conclusions}
In conclusion, we have examined how one can produce a general rotation
of the qubit in  a superposition of the states $|g\rangle$ and $|f\rangle$ by using
the nonadiabatic evolution. Two delayed pulses 
in two-photon resonance have been employed for this
purpose. We have found that each of the parameters - one-photon detuning, 
 amplitudes, phases, and chirpings of the two pulses can provide proper control
 of the qubit rotation. The effect of time separation of the pulses has also
been investigated. We discussed the relevance of two-level approximation in the present context. 
It is clear that logic gate operations
can be implemented as well by using nonadiabatic evolution.    
Finally note that one could preselect a definite time-dependent phase of the 
 pulse to attain a 
desired final state. 

G.S.A. thanks National Science Foundation (grant number INT9605072) for a visit to the University of 
Rochester. He also thanks J. H. Eberly for discussions on the interaction of multilevel 
systems with pulsed fields.

\appendix
\section*{}
We present here an alternative method of orthogonal rotation of a qubit initially in the superposition state $|i\rangle$ [Eq.\ (\ref{initial})]. The state 
orthogonal to $|i\rangle$ is $|k\rangle$  which is given by Eq.\ (\ref{final}).
The method described here for rotation of the qubit from  the state $|i\rangle$
 to $|k\rangle$ is a generalization of the STIRAP process. This is implemented through {\it adiabatic} evolution. 
The Hamiltonian of the system is the same as in (\ref{hamil}), but with $\Delta_i=0$ and $\chi_i=0$ $(i=1,2)$. We rewrite this
in terms of the basis states ($|i\rangle, |k\rangle, |e\rangle$) in the following way : 
\begin{eqnarray}
H&=&\hbar[(\Omega_1 \alpha + \Omega_2 \beta e^{i\phi})|e\rangle\langle i|+(\Omega_1\beta e^{-i\phi}-\Omega_2\alpha)|e\rangle\langle k|\nonumber\\
&&\hspace*{3cm}+\textrm{h.c.}].
\end{eqnarray}
The relevant time-dependent equations for probability amplitudes in the basis
($|i\rangle$, $|k\rangle$, $|e\rangle$) will be 
\begin{eqnarray}
\dot{c}_i&=&-if_1(t)c_e,\nonumber\\
\dot{c}_k&=&-if_2(t)c_e,\nonumber\\
\dot{c}_e&=&-i[f_1^*(t)c_i+f_2^*(t)c_k],
\end{eqnarray}
\label{appeq}
where
\begin{eqnarray} 
f_1(t)&=& \alpha \Omega_1^*(t)+\beta e^{-i\phi}\Omega_2^*(t),\nonumber\\
f_2(t)&=& \beta e^{i\phi}\Omega_1^*(t)-\alpha \Omega_2^*(t).
\end{eqnarray}
\label{f1f2}
We solve these equations under the initial conditions 
\begin{equation}
c_i(-\infty)=1,~~c_k(-\infty)=c_e(-\infty)=0.
\end{equation}
If one chooses the time-dependences of $f_1$ and $f_2$ in the following way 
\begin{eqnarray}
f_1&=& \exp{[-(t-T)^2/\tau^2]},\nonumber\\
f_2&=& \exp{(-t^2/\tau^2)},
\end{eqnarray}
\label{pulseApp}
then, one can transfer the population from $|i\rangle$ to $|k\rangle$ 
adiabatically. This is analogous to the well known STIRAP process in which population is transferred from the state $|g\rangle$ to $|f\rangle$.
Using Eqs.\ (\ref{f1f2}) and (\ref{pulseApp}) we obtain the form of the 
required  pulses $\Omega_1(t)$ and $\Omega_2(t)$    
\begin{eqnarray}
\Omega_1(t)&=& \alpha \exp{[-(t-T)^2/\tau^2]}+\beta e^{-i\phi}\exp{(-t^2/\tau^2)},\nonumber\\
\Omega_2(t)&=& \beta e^{i\phi}\exp{[-(t-T)^2/\tau^2]}-\alpha \exp{(-t^2/\tau^2)}
\end{eqnarray}
to transfer the population from the state $|i\rangle$ to its orthogonal state 
$|k\rangle$.

We thus find that, in the long time limit the population is transferred to the state
$|k\rangle$ through adiabatic process while the level $|e\rangle$ remains almost
unpopulated for all times. Thus the qubit is rotated from $|i\rangle$ to its orthogonal state $|k\rangle$.
One may note that as the evolution described here is analogous to STIRAP, so the robustness of the process
is similar to that of STIRAP.
It is also worth mentioning that one could rotate $|i\rangle$ to any other arbitrary superposition of $|g\rangle$ 
and $|f\rangle$ by just chopping the pulses $\Omega_1(t)$ and $\Omega_2(t)$ at the desired moment. 
But this requires complete control on the pulse area.

\end{multicols}

\end{document}